\documentclass[cameraready]{Interspeech}

\usepackage{amsmath}

\title{A Novel Sentence Stress Detection Framework Leveraging Auxiliary Word-Stress Modeling and Loss Optimization}

\author[affiliation={}, orcid=0009-0006-1825-1918]{Tien-Hong}{Lo}
\author[affiliation={}, orcid=0000-0003-0087-9978]{Fong-Chun}{Tsai}
\author[affiliation={}]{Ting-An}{Hung}
\author[affiliation={}]{Yu-Hsuan}{Hsieh}
\author[affiliation={}, orcid=0000-0002-3538-3756]{Yao-Ting}{Sung}
\author[affiliation={}, orcid=0000-0003-0693-8932]{Berlin}{Chen}

\address{
    National Taiwan Normal University, Taiwan
}

\email{\{teinhonglo,berlin\}@ntnu.edu.tw}

\keywords{Automatic pronunciation assessment, sentence stress detection, word stress detection.}

\usepackage{comment}

\begin{document}

\maketitle

\begin{abstract}
Prosodic stress is a crucial aspect of automatic pronunciation assessment (APA), encompassing both sentence stress detection (SSD) and word stress detection (WSD). SSD highlights semantically salient words that shape discourse meaning, while WSD identifies the primary stressed syllable within each word to ensure lexical clarity. However, most prior work treats SSD and WSD as independent tasks, overlooking their shared reliance on prosodic cues such as pitch, duration, and intensity. To address this gap, we propose an effective SSD approach combining SSD with auxiliary WSD via a novel modeling paradigm. In addition, we introduce a word-span stress regularizer (WSR) that concentrates token-level SSD probabilities within each stressed word span. Experiments on the TinyStress-15K benchmark show that the proposed method outperforms strong baselines, with the complete configuration achieving the best SSD result.
\end{abstract}

\section{Introduction}
\label{sec:intro}

With the rapid advances in computing technology and the growing global population of second-language (L2) learners, automatic pronunciation assessment (APA) has attracted considerable attention and plays an increasingly prominent role in computer-assisted language learning (CALL) \cite{IntroAsaVan2016_hsla}. APA systems are designed to provide timely and objective feedback on learners’ speaking performance, thereby facilitating self-directed improvement and reducing the instructional burden on teachers. Research on APA focuses specifically on evaluating multiple aspects of L2 learners’ speech, such as fluency and prosody \cite{ApaCucchiarini1998_icslp}.

\begin{figure}[t]
  \centering
  \includegraphics[width=0.9\linewidth]{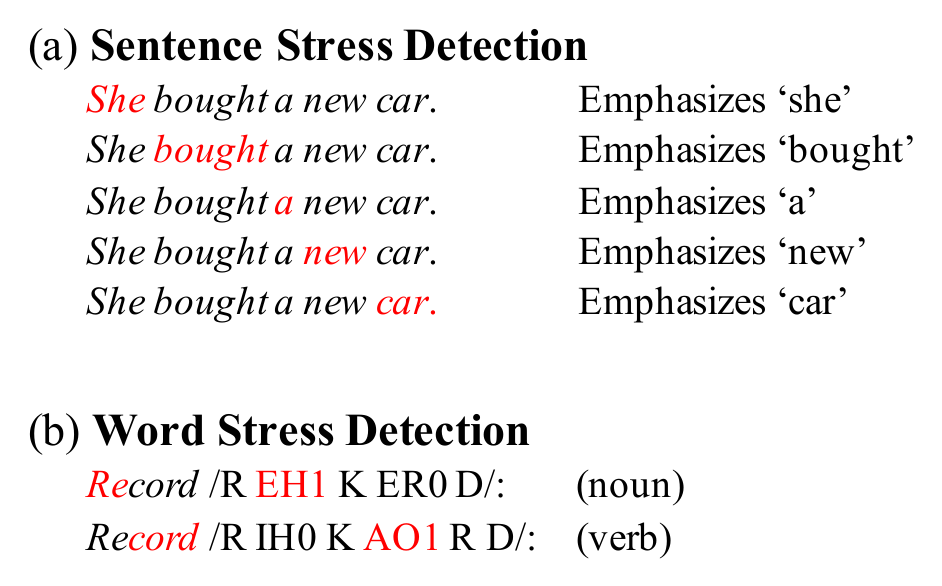}
  \caption{Illustration of stress detection at two linguistic levels. (a) SSD highlights the emphasized word within a sentence by shifting prominence across different lexical items. (b) WSD identifies the primary stressed syllable within a lexical item, where primary stress placement is indicated by the digit “1”.}
  \label{fig:wsd_ssd_illustration}
\end{figure}

Prosodic stress constitutes a critical component of APA, encompassing both sentence stress detection (SSD) and word stress detection (WSD). As displayed in Fig. \ref{fig:wsd_ssd_illustration}, sentence stress detection (SSD) identifies semantically salient words that shape discourse meaning, while word stress detection (WSD) locates the primary stressed syllable within each word to ensure lexical clarity. As for SSD, theoretical studies have framed stress as either a phonological default or a semantic device for emphasis \cite{SsdIntroLadd2008_cambridge,SsdIntroChomsky1968_book,SsdIntroBolinger1972_language}, while conceptual accounts have further characterized its acoustic correlates such as duration, intensity, and pitch \cite{SsdIntroHeuven2018_book}. Building on these acoustic and linguistic foundations, early SSD methodologies primarily employed shallow classifiers alongside handcrafted features \cite{SsdKakouros2016_spcomm,SsdMishra2012_interspeech}, later advancing to neural architectures \cite{SsdMorrison2024_icassp,SsdSeyssel2024_emnlp}. Additionally, some research has demonstrated that exploiting syntactic boundaries and intonation units can further improve prediction \cite{SsdLin2020_interspeech,SsdLee2017_elsevier,SsdSuni2015_arxiv}. More recently, large-scale end-to-end frameworks with stress detection objectives, trained on the synthetic dataset, have achieved competitive benchmark performance \cite{SsdYosha2025_interspeech,SsdHung2025_rocling}.  On the separate front, WSD research has long relied on acoustic cues such as intensity, duration, and pitch \cite{WsdFerrer2015_spcomm,WsdDelmonte1997_eurospeech,WsdTepperman2005_icassp}, later augmented with contextual sonority-based features \cite{WsdYarra2017_icassp} and nucleus-level clustering \cite{WsdDeshmukh2009_spcomm}. Early classifiers included decision trees and SVMs \cite{WsdJohnson2015_ijst}, while deep models such as CNNs and transformers have since advanced word stress classification \cite{WsdShahin2016_interspeech,WsdRuan2019_arxiv,WsdMallela2023_iscs}. Recent work has instead integrated adaptive loss functions that embed linguistic constraints directly into training, reducing the need for post-hoc correction \cite{WsdAluru2025_icassp}. More broadly, structured training objectives for prosody often take the form of soft constraints or regularizers, such as entropy penalties or sparsity-inducing terms, to encourage well-formed stress patterns. Despite the continued efforts, SSD and WSD continue to be modeled independently, leaving unexplored their shared reliance on prosodic cues and the potential benefits of a unified framework.

To tackle the aforementioned challenge, we propose \textbf{STRAW}, a \textbf{S}entence s\textbf{T}ress detection (SSD) framework built on a frozen Whisper backbone that augments token-level SSD with a word-span stress \textbf{R}egularizer (WSR) and an \textbf{A}uxiliary phone-level \textbf{W}ord stress detection (WSD) branch. Unlike previous approaches that treat SSD and WSD in isolation, the proposed framework supports both tasks within a common frozen-Whisper framework. Experiments on the TinyStress \cite{SsdYosha2025_interspeech} benchmark dataset demonstrate improvements over strong baselines. The main contributions of this work are summarized as follows:
\begin{enumerate}
    \item We present a unified framework for stress detection with separate task-specific branches for Sentence Stress Detection (SSD) and Word Stress Detection (WSD).

    \item We introduce a word-span stress regularizer (WSR), a linguistically motivated constraint that discourages diffuse SSD probabilities across the subword tokens of a stressed word.
\end{enumerate}



\begin{figure}[t]
  \centering
  \includegraphics[width=\linewidth]{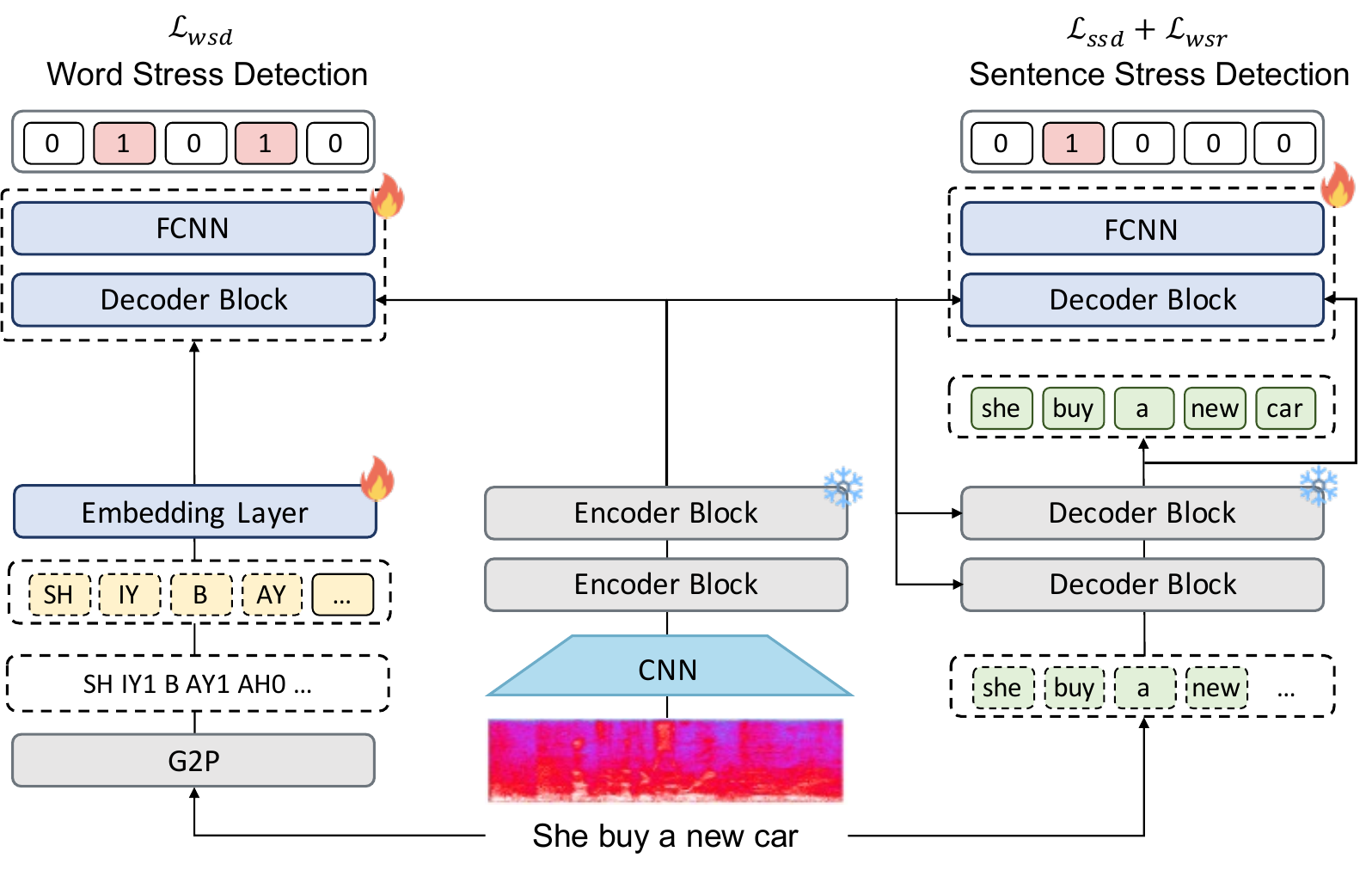}
  \caption{Overall architecture of the proposed framework. A frozen Whisper backbone provides encoder representations to both heads and decoder representations to the SSD head. Sentence stress detection (SSD) operates at the token level with an additional word-span stress regularizer (WSR), while word stress detection (WSD) operates at the phone level. The SSD and WSD heads are task-specific and do not directly exchange hidden states or predictions.}
  \label{fig:framework}
\end{figure}

\section{Method}
\label{sec:method}

In this section, prosodic stress detection is formulated as a supervised classification task over spoken utterances. As illustrated in Figure~\ref{fig:framework}, each utterance is represented as a log-mel spectrogram $\mathbf{s}_{1:T}$ with $T$ acoustic frames. 

The corresponding transcription is expressed as a word sequence $\mathbf{w}_{1:L}$, a token sequence $\mathbf{t}_{1:M}$, and a phone sequence $\mathbf{p}_{1:N}$ using an off-the-shelf grapheme-to-phoneme (G2P) converter with CMU-style stress-marked phones. Subsequently, the Whisper backbone is employed to encode $\mathbf{s}$ through its encoder and decoder blocks, producing contextualized encoder states $\mathbf{e}_{1:T}$ and decoder states $\mathbf{d}_{1:M}$. Two task-specific modules, sentence stress detection (SSD) and word stress detection (WSD), are introduced on top of the frozen Whisper backbone. Specifically, SSD is described in Section~\ref{subsec:ssd}, WSD is described in Section~\ref{subsec:wsd}, and the word-span stress regularizer (WSR) is described in Section~\ref{subsec:wsr}.

In our experiments, the Whisper backbone is frozen and we train only the task-specific heads and the phone embedding layer. SSD is trained at the token level, where each token receives a binary stress label $y^{ssd}_{1:M} \in \{0,1\}$, while evaluation aggregates predictions $\hat{y}^{ssd}_{1:M}$ to the word level by assigning a word as stressed if any of its tokens is predicted as stressed. WSD, in contrast, operates at the phone level, where each phone is labeled $y^{wsd}_{1:N} \in \{0,1\}$ as stressed or unstressed.

\subsection{Sentence Stress Detection}
\label{subsec:ssd}
As shown in the right branch of Figure~\ref{fig:framework}, the SSD module operates at the token level. For each utterance, the Whisper decoder-state sequence $\mathbf{d}_{1:M}$ serves as the queries, while the encoder-state sequence $\mathbf{e}_{1:T}$ serves as the keys and values. A Transformer decoder block integrates the two sources of information, followed by a fully connected neural network (FCNN) to predict token-level stress:
\begin{align}
\mathbf{d}^{ssd}_{1:M} &= \text{Transformer}^{ssd}(\mathbf{d}_{1:M}, \mathbf{e}_{1:T}), \\
\hat{\mathbf{y}}^{ssd}_{1:M} &= \mathrm{Softmax}\left(\mathrm{FCNN}^{ssd}(\mathbf{d}^{ssd}_{1:M})\right),
\end{align}

where $\hat{\mathbf{y}}^{ssd}_m \in \mathbb{R}^{2}$ is a two-class posterior over \texttt{unstressed} vs.\ \texttt{stressed} for the $m$-th token, and we denote $\hat{y}^{ssd}_m$ as the stressed-class probability. The loss function of $\mathcal{L}_{SSD}$ is computed using cross-entropy over token-level SSD labels.

\subsection{Word Stress Detection}
\label{subsec:wsd}
The left branch of Figure~\ref{fig:framework} shows the phone-level WSD module. Given an utterance-level phone sequence $p_{1:N}$, we obtain trainable phone embeddings $\mathbf{p}_{1:N}$ through an embedding layer. We then compute phone-aware acoustic representations via cross-attention, where $\mathbf{p}_{1:N}$ serves as queries and the Whisper encoder-state sequence $\mathbf{e}_{1:T}$ serves as the keys and values. The resulting phone representations are then passed to an FCNN to yield phone-level stress predictions:
\begin{align}
\mathbf{d}^{wsd}_{1:N} &= \text{Transformer}^{wsd}(\mathbf{p}_{1:N}, \mathbf{e}_{1:T}), \\
\hat{\mathbf{y}}^{wsd}_{1:N} &= \mathrm{Softmax}\left(\mathrm{FCNN}^{wsd}(\mathbf{d}^{wsd}_{1:N})\right),
\end{align}

where $\hat{\mathbf{y}}^{wsd}_n \in \mathbb{R}^{2}$ is a two-class posterior over \texttt{unstressed} vs.\ \texttt{stressed} for the $n$-th phone, and we denote $\hat{y}^{wsd}_n$ as the stressed-class probability. The WSD loss $\mathcal{L}_{WSD}$ is optimized using the cross-entropy objective over phone-level WSD labels.

\subsection{Word-Span Stress Regularizer}
\label{subsec:wsr}

Since the SSD branch is formulated at the token level, a lexical word may correspond to a span of subword tokens under Whisper tokenization. When a word is annotated as stressed in SSD, the token-level SSD head may assign elevated stress probabilities to multiple tokens within its subword span, yielding an underconstrained within-word allocation of prominence. We therefore encourage a single dominant token-level stress position within each ground-truth stressed word span. Accordingly, we introduce a word-span stress regularizer over token-level SSD outputs to encourage a unique dominant stressed position for each ground-truth stressed word $w_m$ with token span $[b_m,e_m]$:
\begin{align}
i^{\star} &= \mathop{\arg\max}_{i \in \{b_m,\ldots,e_m\}} \hat{y}^{ssd}_i, \\
\omega(w_m) &= \Big|1 - \sum_{i=b_m}^{e_m} \hat{y}^{ssd}_i\Big|,
\end{align}
where $\omega(w_m)$ measures the deviation of the total SSD stress probability within $w_m$ from one and weights the subsequent peak-sharpening term. The word-span stress regularizer $\mathcal{L}_{WSR}$ for $w_m$ is formulated as:
\begin{multline}
\mathcal{L}_{WSR}(w_m) = \omega(w_m) \cdot \Big( 
- \log \hat{y}^{ssd}_{i^\star} \\
- \sum_{i \in \{b_m,\ldots,e_m\}\setminus\{i^\star\}}
    \log \big(1-\hat{y}^{ssd}_i\big) \Big),
\end{multline}

The final $\mathcal{L}_{WSR}$ is averaged over the ground-truth stressed words in the utterance, encouraging the dominant SSD probability $\hat{y}^{ssd}_{i^\star}$ to approach $1$ while suppressing the remaining positions within the same word span. $\mathcal{L}_{WSR}$ is used as a regularization constraint over the SSD posteriors.

In this view, $\mathcal{L}_{SSD}$ provides coarse supervision for identifying stressed words, while $\mathcal{L}_{WSR}$ serves as a within-word sharpening constraint that resolves the allocation ambiguity induced by subword tokenization.

\subsection{Optimization}
\label{subsec:mtl}
The final loss function is formulated as:
\begin{align}
\label{eq:final_loss}
\mathcal{L} &= \alpha \cdot \mathcal{L}_{SSD}
            + \beta \cdot \mathcal{L}_{WSD}
            + \lambda \cdot \mathcal{L}_{WSR},
\end{align}
where $\alpha, \beta, \lambda \geq 0$ are the weighting coefficients. This design enables the framework to capture SSD, WSD, and WSR in a unified manner.

\begin{table}[]
\centering
\caption{Statistics of the TinyStress-15K corpus.}
\label{tab:dataset}
\begin{tabular}{l|c|c|l|c}
\hline
 & \#Audios & \#Tokens & \#Stress & \#Unstress \\ \hline
Train & 13,500   & 171,206  & 22,537   & 148,669     \\ \hline
Valid & 1,500    & 18,676   & 2,541    & 16,135      \\ \hline
Test  & 1,000    & 12,548   & 1,705    & 10,843      \\ \hline
\end{tabular}
\end{table}

\section{Experiment Setup}
\label{sec:experiments_setup}

\subsection{Dataset}
\label{sec:dataset}

Experiments are conducted on the TinyStress-15K corpus \cite{SsdYosha2025_interspeech}, which comprises synthetic speech with word-level stress annotations. Each instance includes an input utterance $\mathbf{a}$, token-level sentence stress detection (SSD) labels $y^{ssd}_m$ derived via Whisper tokenization, and phone-level word stress detection (WSD) labels $y^{wsd}_n$ obtained using a grapheme-to-phoneme (G2P) toolkit\footnote{\url{https://github.com/Kyubyong/g2p.git}}. We apply the G2P converter to the complete reference transcript rather than to isolated words. We derive $y^{wsd}_n$ from the G2P stress digits by labeling only primary-stressed phones (digit ‘1’) as 1, and mapping all other phones, including digit ‘2’, to 0. As summarized in Table~\ref{tab:dataset}, the training split contains 13.5k utterances and 171k tokens, including approximately 22.5k stressed and 148k unstressed tokens. The validation split comprises 1.5k utterances with 18.6k tokens, and the test split comprises 1k utterances with 12.5k tokens. Following the split protocol of \cite{SsdYosha2025_interspeech}, we adopt the official training and test partitions, and hold out 10\% of the training split as a validation set. This design preserves a consistent evaluation protocol while providing a held-out validation set for model selection.

\subsection{Comparison Methods}
\label{subsec:baselines}

We compare the proposed framework against both alignment-based and alignment-free baselines. The alignment-based systems use ground-truth word-level timestamps or the Montreal Forced Aligner (MFA) to extract prosodic features such as duration, energy, and pitch, followed by a BLSTM classifier \cite{SsdLin2020_interspeech}. These models represent traditional pipelines that heavily rely on accurate alignment. The alignment-free baseline is WhiStress \cite{SsdYosha2025_interspeech}, which extends Whisper with a stress detection head trained on synthetic resources. WhiStress reports strong performance without requiring timestamps, but it treats sentence stress as an isolated objective and does not impose lexical stress constraints. Our extensions build upon this alignment-free setting by incorporating an auxiliary WSD objective and a WSR, enabling sentence- and word-stress prediction within a unified framework.

\subsection{Implementation Details}
\label{sec:impl}

Model configurations were initialized using pretrained models from the HuggingFace Transformers library \cite{ToolHugefaceWolf2020_emnlp}. We employed whisper-small\footnote{\url{https://huggingface.co/openai/whisper-small}} as the backbone model and extracted hidden states from the ninth layer of the Whisper encoder/decoder for downstream modeling, following the configuration of WhiStress \cite{SsdYosha2025_interspeech}. We freeze all Whisper parameters and train only the SSD head, the WSD head, and the phone embedding layer. Both prediction heads were optimized in the same training run using the multi-task objective in Section~\ref{sec:method}. Unless otherwise stated, we set $\alpha=1.0$, $\beta=1.0$, and $\lambda=1.0$ in Eq. \ref{eq:final_loss} for all experiments. All models were trained on an NVIDIA 3090 GPU using the AdamW optimizer with a batch size of 16 and an initial learning rate of 1e-4. All models were trained for 20 epochs, and the checkpoint achieving the highest binary F1 score for SSD on the validation set was selected for evaluation. At inference time, SSD is performed given the input speech and a Whisper token sequence, while WSD estimates phone-level stress based on a G2P-derived phone sequence. In our \textit{read-aloud} APA setting, the reference transcript provides Whisper decoder tokens for SSD and a G2P-derived phone sequence for WSD, without requiring an additional ASR pass. The WSD branch is optional at inference and may be disabled when only SSD feedback is required.

\subsection{Evaluation Metrics}
\label{sec:metric}

Evaluation adheres to the protocol in \cite{SsdYosha2025_interspeech}. We report precision, recall, and F1 with respect to the stressed class. For SSD, token-level predictions are mapped back to their corresponding word spans in the reference transcript, and a word is marked as stressed if any of its constituent tokens is predicted as stressed. For WSD, predictions are produced at the phone level, and precision, recall, and F1 are computed directly over stressed phone labels.

\section{Experimental Results}
\label{sec:experiment_results}

\begin{table}[]
\centering
\caption{SSD performance on TinyStress-15K. STRAW denotes our proposed framework with auxiliary WSD and the WSR. Ablations remove the corresponding components.}
\label{tab:baseline}
\begin{tabular}{lccc}
\hline
\textbf{Model}                         & \textbf{Precision} & \textbf{Recall} & \textbf{F1}    \\ \hline
GT alignment \cite{SsdYosha2025_interspeech} & 0.862              & 0.853           & 0.858          \\
MFA \cite{SsdYosha2025_interspeech}          & 0.776              & 0.859           & 0.815          \\
WhiStress \cite{SsdYosha2025_interspeech}    & 0.912              & 0.906           & 0.909          \\ \hline
STRAW                                  & \textbf{0.945}              & \textbf{0.924}           & \textbf{0.934}          \\
- WSD                                  & 0.938              & 0.906           & 0.922          \\
- WSR                                  & 0.942              & 0.917           & 0.929          \\
- WSR \& WSD                           & 0.940     & 0.892  & 0.915 \\ \hline
\end{tabular}%
\end{table}

\begin{table}[]
\centering
\caption{WSD performance of STRAW on TinyStress-15K.}
\label{tab:wsd_performance}
\begin{tabular}{lccc}
\hline
\textbf{Model}                         & \textbf{Precision} & \textbf{Recall} & \textbf{F1}    \\ \hline
STRAW                                  & 0.924              & 0.916           & 0.920         \\
- WSR                                  & 0.935              & 0.907           & 0.921          \\
\hline
\end{tabular}%
\end{table}

\begin{figure*}[!t]
  \centering
  \begin{minipage}[b]{.48\linewidth}
    \includegraphics[width=\linewidth]{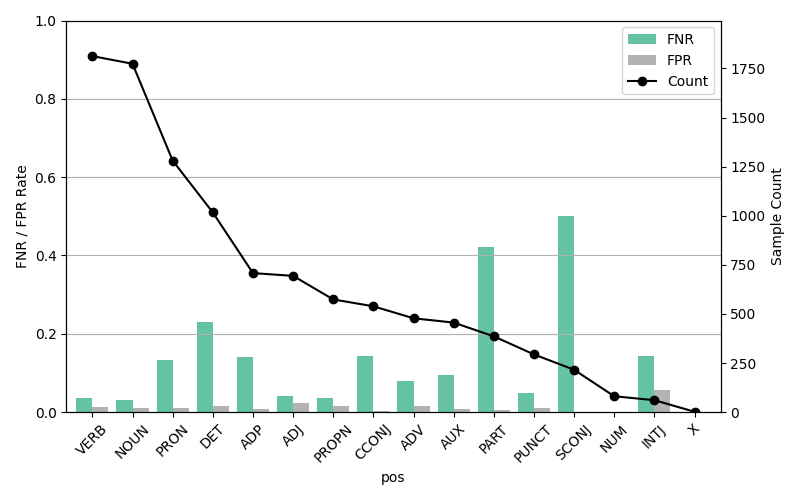}
    \vspace{2pt}\centerline{\small (a) WhiStress}
  \end{minipage}\hfill
  \begin{minipage}[b]{.48\linewidth}
    \includegraphics[width=\linewidth]{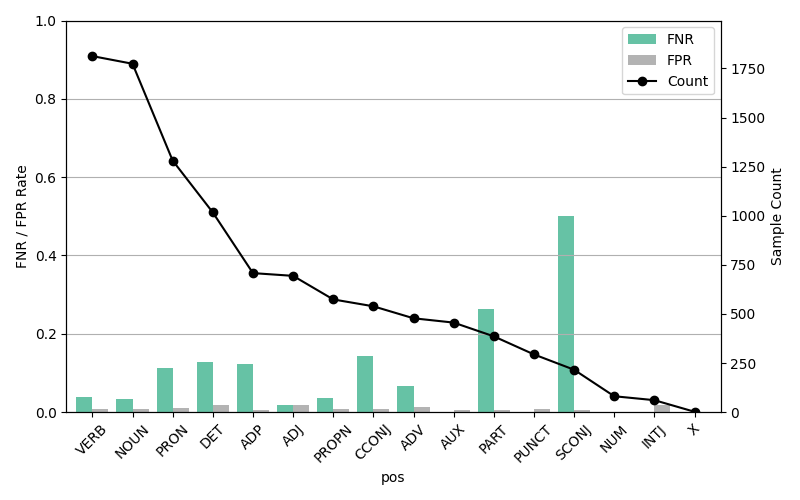}
    \vspace{2pt}\centerline{\small (b) STRAW}
  \end{minipage}
  \caption{Error analysis by part-of-speech (POS) on the test set. Bars show false negative rate (FNR) and false positive rate (FPR). The black line indicates the sample count per POS.}
  \label{fig:pos_error}
\end{figure*}

\subsection{SSD Performance Comparison with Baselines}
\label{subsec:baseline}

Table~\ref{tab:baseline} summarizes SSD results on TinyStress-15K. We include both alignment-based and alignment-free baselines for reference; overall, alignment-based pipelines are sensitive to boundary quality, whereas the alignment-free WhiStress baseline provides a strong Whisper-based point of comparison. Building on the same frozen-Whisper setting, our proposed STRAW achieves the best overall performance, reaching an F1 of 0.934 and improving over WhiStress by 2.5 points (0.909 $\rightarrow$ 0.934) with consistent gains in both precision and recall.

The ablation results show that removing WSD, WSR, or both reduces F1 to 0.922, 0.929, and 0.915, respectively. Because the two heads have separate trainable parameters, the WSD ablation does not establish direct knowledge transfer from WSD to SSD. The WSR ablation is consistent with its intended role of reducing diffuse SSD probabilities within stressed word spans. Overall, the complete configuration achieves the highest score among the evaluated settings.

\subsection{WSD Performance}
\label{subsec:wsd_exp}

Table~\ref{tab:wsd_performance} reports WSD performance on TinyStress-15K. As TinyStress-15K is primarily used for SSD benchmarking, directly comparable external WSD baselines are limited; we report WSD results to contextualize the auxiliary task behavior. Overall, WSD achieves consistently strong results, with F1 around 0.92 across settings, suggesting that phone-level stress prediction on this benchmark is relatively well-conditioned under the provided supervision. In particular, STRAW attains 0.924 precision, 0.916 recall, and 0.920 F1. We further examine whether WSR affects WSD. Removing WSR yields only a marginal change in F1 (0.920 $\rightarrow$ 0.921). This behavior is expected given our training design, in which SSD is prioritized and WSR is introduced primarily to shape stress distributions for word-level prominence, whereas WSD is optimized with its own supervised objective. A promising direction for future work is to introduce explicit coupling between SSD and WSD, for example through shared trainable modules or cross-task consistency constraints.

\subsection{Error Analysis}
\label{subsec:pos}

To better understand the behavior of the proposed models, we analyze error distribution across part-of-speech (POS) categories, as illustrated in Figure~\ref{fig:pos_error}. The baseline system exhibits high false negative rates on functional categories, particularly PART (particle) and DET (determiner). The complete STRAW configuration shows a modest reduction in these errors, with additional improvements observed for AUX (auxiliary) and ADP (adposition). False positives remain consistently low, indicating that precision is not compromised. One exception is SCONJ (subordinating conjunction)\footnote{For the full POS set, please refer to \url{https://universaldependencies.org/u/pos/}}, where high false negative rates persist due to the limited number of samples. These observations suggest that the proposed framework primarily benefits frequent categories with sufficient coverage, which aligns with the overall recall gains reported in Table \ref{tab:baseline}.

\section{Conclusion}
\label{sec:conclusion}

This work introduces STRAW, an SSD framework that augments token-level SSD with an auxiliary WSD head and a WSR. By supporting sentence- and word-level stress analysis within a unified framework and explicitly constraining token-level SSD predictions within stressed word spans, the proposed design mitigates within-word ambiguity induced by subword tokenization. Experiments on TinyStress-15K demonstrate improvements over strong baselines, with the complete configuration achieving the highest F1 score among the evaluated configurations. Three limitations remain. First, the current word-stress detection simplifies lexical stress to a single primary position per word and does not explicitly model syllabic structure or secondary stress. Second, the frozen backbone and separate task-specific heads do not provide direct interaction between SSD and WSD. Third, the present evaluation is confined to synthetic speech due to the scarcity of large-scale, fine-grained stress annotations. Future work will address these issues by extending STRAW to spontaneous and human-recorded speech (e.g., Aix-MARSEC \cite{Auran2004TheAP}, Expresso \cite{nguyen2023expresso}, and EmphAssess \cite{de2023emphassess}), introducing explicit coupling between the SSD and WSD heads, and incorporating richer linguistic knowledge to better capture prosodic structure in SSD.

\newpage

\section{Generative AI Use Disclosure}
ChatGPT was used for language editing and for rephrasing selected passages to improve clarity and coherence. The authors take full responsibility for the technical content and claims, and they verified all descriptions of methods, experiments, and results.

\bibliographystyle{IEEEtran}
\bibliography{mybib}

\end{document}